\documentclass[11pt,twoside,usenatbyb]{article}
\usepackage{asp2010}

\resetcounters

\begin{document}

\title{SXP 1062: an Evolved Magnetar in a BeXB ?}
\author{R. Turolla$^1$ and S.B. Popov$^2$}
\affil{Department of Physics and Astronomy, University of Padova,
Italy$^1$}
\affil{Sternberg Astronomical Institute, Lomonosov Moscow State
University, Russia$^2$}

\begin{abstract}
SXP 1062, a newly discovered Be/X-ray binary in the Small Magellanic
Cloud, provides the first example of a robust association
with a supernova remnant (SNR). The short age estimated for the
SNR qualifies SXP 1062 as the youngest known source in its class,
$\tau\approx 10^4\ \textrm{yr}$. Here we discuss possible evolutionary
scenarios for SXP 1062 in the attempt to reconcile its long
spin period, $P=1062\ \textrm{s}$, and short age. Our results indicate that
SXP 1062
may host a neutron star born with a large initial magnetic
field, typically in excess of $\sim 10^{14}\, {\rm G}$, which then
decayed to $\sim 10^{13}\, {\rm G}$.
\end{abstract}

\section{Introduction}
Be/X-ray binaries (BeXBs) are the largest subclass of high-mass X-ray
binaries (HMXBs)
and are both transient and
persistent X-ray sources.
The latter  exhibit a rather flat lightcurve,
lower X-ray luminosity ($L\sim 10^{34}$--$10^{35}$ erg/s), longer
spin and orbital periods ($P\ga 200$ s, $P_{orb}\ga 200$ d; e.g.
\citealt{reig11}).

Very recently \cite{henbru11} and \cite{hab11}
reported the discovery of a new BeXB in the
Small Magellanic Cloud (SMC). While SXP 1062 has the typical
properties of a persistent BeXB
($L\sim 6\times 10^{35}$ erg/s and $P\sim 1062$ s) what makes this source
unique is its robust association with a supernova remnant (SNR).
This allows for the first time to estimate the NS age in a BeXB,
$\tau\sim 2$--$4\times 10^4$ yr.

According to the standard
picture, there are four stages in the spin evolution of a NS:
ejector, propeller, accretor, and georotator \cite[e.g.][]{lip92}.
Once the NS enters the accretor stage,
its spin period quickly settles at an equilibrium
value, $P_{eq}$. Ultra-long periods ($P_{eq} > 1000$ s) can be reached
without invoking super-strong fields ($B>10^{14}$ G) if a subsonic
propeller stage sets in \citep{iks07}, or a settling regime forms in
the
accretion flow \citep{shak11}.

The previous argument
implicitly relies on the assumption that the present age of the
source is long enough for the NS to have entered the propeller
stage.
Normally, a NS in a
BeXB with $B>10^{12}$ G starts its evolution in the ejector
phase. Its duration is $\sim 10^6 (B/10^{12}{\rm
G})^{-1}(\dot M/10^{15}{\rm g/s})^{-1/2}$ yr. This is comfortably
below
the lifetime of the Be companion.
In the case of SXP 1062, however,
it would be impossible for the NS to enter the propeller stage
in a time as short as a few
$\times 10^4$ yr, the estimated SNR age, for typical values of $B$
and $\dot M$. The accretion rate in SXP 1062 is $\dot M=L/\eta
c^2\sim 6\times 10^{15}$~g/s for an efficiency $\eta=0.1$, so this
points to a highly magnetized NS, with an initial magnetic field
substantially above $10^{12}$ G.
Here we summarize the results presented in \cite{pt12} and add some new
considerations.

\section{Spin Evolution in SXP 1062}


In the ejector phase the light cylinder
radius, $R_{lc}$,
is typically
smaller than the gravitational capture radius, $R_G=2GM/V^2$,
where $V$ is the velocity of
matter far from the star ($M_{NS}=1.4 M_\odot$,
$R_{NS}=10$ km and moment of inertia $I=10^{45}\ \mathrm{g\,cm}^2$ are
assumed henceforth). The transition to the propeller stage
occurs when the ram pressure, $P_{dyn}=\rho V^{2}/2$, balances the
outgoing flux of electromagnetic waves and relativistic particles
$P_{PSR}=\dot E/(4\pi R^2c)$, at $R_G$ ($\dot E$ is the rotational
energy loss rate of the pulsar).
The critical period for the
transition follows by requiring that
$P_{dyn}(R_{G})=P_{PSR}(R_{G})$, together with the standard
expression for magneto-dipole losses and mass conservation,

\begin{equation}
P_{ej}=2 \pi \left( \frac43 \frac{B^2R_{NS}^6}{\dot M V
c^4}\right)^{1/4} \sim 0.31 \, V_{300}^{-1/4} \dot
M_{16}^{-1/4} B_{12}^{1/2}\,{\mathrm s}.
\end{equation}

From the magneto-dipole formula, assuming constant $B$ and very
short initial period, it follows that

\begin{equation}
\tau_{ej} =\frac{3Ic^3 P_{ej}^2}{16\pi^2 B^2 R_{NS}^6}\sim 1.5\,
\dot M_{16}^{-1/2}  V_{300}^{-1/2} B_{12}^{-1}\, {\rm Myr}\,.
\end{equation}
The dipole field in a wind-fed NS has been estimated by \cite{shak11}
under the assumption that the star is spinning at the
equilibrium period
\begin{equation}
\label{bshak}
B_{12} \sim 8.1\, \dot M_{16}^{1/3}
V_{300}^{-11/3}\left(\frac{P_{1000}}{P_{orb\,
300}}\right)^{11/12}\, {\mathrm G}.
\end{equation}
This gives  $\tau_{ej}\sim 0.2$ Myr for SXP 1062,
a factor 10 longer than the SNR age.

Within this framework, an obvious possibility to shorten the
ejector phase in SXP 1062 is to invoke a higher dipole field.
However, if the present field is that given by the previous
expression this implies that $B$ must have been stronger in the
past and then decayed to its present value
\cite[see e.g.][]{pons09, popov10, turolla11}.

The period evolution in the ejector
stage is governed by magneto-dipolar losses and we adopted the simplified
model of  \cite{aguil08} for the evolution of $B$.
The NS enters the
propeller phase as soon as the dynamical pressure exerted by the
incoming material overwhelms the pulsar momentum flux at the
gravitational radius.
Spin-down is expected to be very efficient in the propeller phase, so
its duration is quite short.
Finally, to follow the period evolution in the accretor stage
we assume the settling accretion regime recently proposed by
\cite{shak11}.

\begin{figure}
\begin{center}
\includegraphics[angle=0,height=7.truecm]{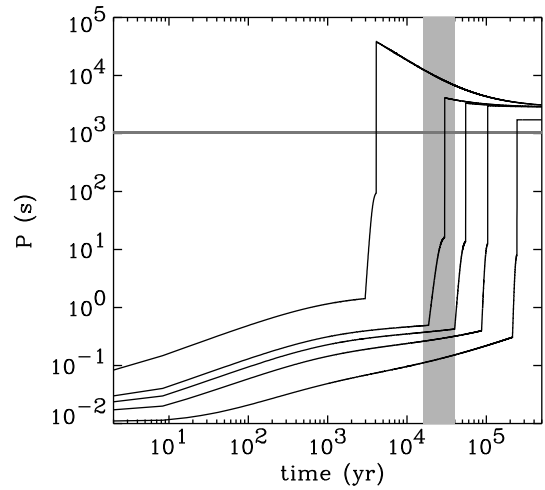}
\end{center}
\caption{\label{fig1} The spin period evolution for
$B_0=4\times
10^{14}$, $10^{14}$, $7\times 10^{13}$, $4\times 10^{13}$,
$10^{13}\, {\rm G}$ (solid lines, from top to bottom). The shaded
areas mark the age and period
of SXP 1062 with the respective uncertainties. }
\end{figure}

We solved numerically the equation for the period evolution in the
three stages starting from $t_0=0.01$ s with an initial period
$P_0=0.01$ s. The accretion rate was fixed to $\dot M=6\times
10^{15}$ g/s, together with $P_{orb}=300$ d, $V=300$ km/s,
Ohmic decay timescale $\tau_O=10^6$ yr,
and relic field $8\times 10^{12}$ G.
Figure \ref{fig1}
illustrates the results for typical runs
with different values of the initial field $B_0$.
The main result is that a quite large initial field is
required in order for SXP 1062 to enter the propeller phase (and
quickly start accreting) in a time as short as a few $\times 10^4$
yr.
For the case at hand it has to be $B_0>10^{14}$ G for this
to occur. However, the result is not very sensitive to the actual choice
of the Hall decay timescale, $\tau_H$, and angle between spin and magnetic
axis, $\alpha$.
The conclusion that SXP 1062 harbours an
initially strongly magnetized NS seems therefore quite robust.


\section{Discussion}

Normally, one expects
that accreting X-ray pulsars spin close to their equilibrium
period. However, for a young system like SXP 1062 this appears far from
granted. \cite{hab11}
reported a large spin-down rate for
SXP 1062 ($\dot P \sim 95$ s/yr) which may suggest that $P_{eq}$
has not been reached yet. According to the standard evolutionary
scenario \citep{lip92},
the (maximum) spin-down rate in the
accretor stage is $\dot P \sim 2\pi B^2R_{NS}^6/(GMI)$ which implies $B
\sim 3\times 10^{14}$ G for $\dot P \sim 100$ s/yr. On the other
hand, if Shakura et al. spin-down formula is used to estimate the
magnetic field, a much lower value is obtained, $B \sim 10^{13}$
G, very close to what is predicted assuming that the source spins
at the equilibrium period. This supports a picture in which the NS
actually rotates close to $P_{eq}$. A further argument in favor of
this is the very short duration of the spin-down phase in the
accretor stage, which makes it very unlikely to catch the source
in this state. Our conclusion is that both the young age and the
large spin-down rate of SXP 1062 argue in favor of an initially
highly magnetic NS which experienced field decay.

Alternative scenarios to explain the long period and short age of
the source can be envisaged. For instance, \cite{hab11}
suggested that the NS could have been born with an initial period
$\gg 0.01$ s. The value of $P_0$ can be
evaluated by requiring that the end of the ejector stage is
reached in less than the source age, and it turns out to be $\sim
1$ s for $B \sim 10^{13}$ G. 
If this is the case no field decay is
required.
Another possibility
is that the NS in could have been surrounded by a debris disc, which
could also lead to rapid spin-down and large period, as suggested
for the enigmatic source RCW 103 \citep{deluca06,li07} (see also a
recent e-print by \citet{yan}).
Although this remains a
possibility worth of further investigations, preliminary
calculations indicate that the disc has be be quite massive
($\ga 10^{-2}M_\odot$) for this to work with a  field $\sim
10^{13}$~G.

If SXP 1062 indeed contains an initially strongly magnetized
neutron star, then studies of this system can shed light on the
origin of magnetars.
\cite{chashkina11}
have recently derived estimates of the
B-field in HMXBs using Shakura et al. model and no evidence of
ultra-high fields was found.
A very recent, quite robust case for a possible magnetar in a binary
system was made by \cite{reig12} (see also Reig et al., this
proceedings). According to eq. (\ref{bshak}), in fact, the field of the NS
in the BeXB 4U 2206+54
($P\sim 5560$~s, $\dot M\sim 3\,\times 10^{15}$~g/s, $V\sim
350$~km/s, $P_{orb}\sim 10$~d) is about $4 \times 10^{14}$~G.
The original estimate in \cite{reig12} also provides similar values.
\cite{bp09} studied several options to produce a
rapidly rotating stellar core just before the collapse, so that the dynamo
mechanism can operate. This turns out to be possible, due to
tidal synchronization, in a close binary with an orbital period
$\la 10$ d. Using the on-line tool for binary evolution
\cite[http://xray.sai.msu.ru/sciwork/scenario.html;][]{lpp96}, we
find that a binary with initial separation $30\, R_\odot$ and masses 27 and
9~$M_\odot$ can match the requirements. Before the SN explosion the
orbital period is about 2 d, and becomes $\la 10$~d
after (the value and direction of the kick is
important in fixing the orbital period). Other combinations of parameters
are possible, too. What is
noticeable is that in such a system tidal synchronization can result in a
rapidly rotating core which later produces a magnetar.

Despite magnetars were searched for in binary systems long since, only
very recently promising candidates have been proposed.
Further studies of magnetar candidates in binaries will be crucial
in shedding light on the origin of these peculiar objects.

\acknowledgements
RT is
partially supported by INAF under a PRIN 2011 scheme and SBP through
RFBR grant 10-02-00059. We thank Pablo Reig for discussions and the
organizers in Zielona Gora for this successful meeting.


\end{document}